\documentclass[sigconf,authorversion]{acmart}

\usepackage{listings} 
\usepackage{xcolor}
\definecolor{codered}{rgb}{0.6,0,0}
\definecolor{codegreen}{rgb}{0,0.6,0}
\definecolor{codegray}{rgb}{0.5,0.5,0.5}
\definecolor{codepurple}{rgb}{0.58,0,0.82}
\definecolor{backcolour}{rgb}{0.95,0.95,0.92}
\lstdefinestyle{mystyle}{
    backgroundcolor=\color{backcolour},   
    commentstyle=\color{codegreen},
    keywordstyle=\color{magenta},
    numberstyle=\tiny\color{codegray},
    stringstyle=\color{codepurple},
    basicstyle=\ttfamily\footnotesize,
    breakatwhitespace=false,         
    breaklines=true,                 
    captionpos=b,                    
    keepspaces=true,                 
    numbersep=5pt,                  
    showspaces=false,                
    showstringspaces=false,
    showtabs=false,                  
    tabsize=2
}
\usepackage{enumitem}

\AtBeginDocument{%
  \providecommand\BibTeX{{%
    \normalfont B\kern-0.5em{\scshape i\kern-0.25em b}\kern-0.8em\TeX}}}

\setcopyright{acmcopyright}
\copyrightyear{2023}
\acmYear{2023}
\acmDOI{}

\acmConference[SIGCSE '23]{SIGCSE '23: ACM Technical Symposium on Computer Science Education}{March 15--18, 2023}{Toronto, Ontario, Canada}
\acmBooktitle{SIGCSE '23: ACM Technical Symposium on Computer Science Education,
  March 15--18, 2023, Toronto, Ontario, Canada}
\acmPrice{15.00}
\acmISBN{978-1-4503-XXXX-X/18/06}

\begin{document}



\title[Large Language Model Explanations of Code]{Experiences from Using Code Explanations Generated by Large Language Models in a Web Software Development E-Book}



\author{Stephen MacNeil}
\affiliation{%
  \institution{Temple University}
  \city{Philadelphia}
  \state{PA}
  \country{USA}
  \postcode{19122}
}
\email{stephen.macneil@temple.edu}
\orcid{0000-0003-2781-6619}

\author{Andrew Tran}
\affiliation{%
  \institution{Temple University}
  \streetaddress{1801 N Broad St}
  \city{Philadelphia}
  \state{PA}
  \country{USA}
  \postcode{19122}
}
\email{andrew.tran10@temple.edu}

\author{Arto Hellas}
\affiliation{%
  \institution{Aalto University}
  \city{Espoo}
  \country{Finland}}
\email{arto.hellas@aalto.fi}
\orcid{0000-0001-6502-209X}

\author{Joanne Kim}
\affiliation{%
  \institution{Temple University}
  \city{Philadelphia}
  \state{PA}
  \country{USA}}
\email{joanne.kim@temple.edu}
\orcid{0000-0001-7646-2373}

\author{Sami Sarsa}
\affiliation{%
  \institution{Aalto University}
  \city{Espoo}
  \country{Finland}}
\email{sami.sarsa@aalto.fi}
\orcid{0000-0002-7277-9282}

\author{Paul Denny}
\affiliation{%
  \institution{The University of Auckland}
  \city{Auckland}
  \country{New Zealand}}
\email{paul@cs.auckland.ac.nz}
\orcid{0000-0002-5150-9806}

\author{Seth Bernstein}
\affiliation{%
  \institution{Temple University}
  \streetaddress{1801 N Broad St}
  \city{Philadelphia}
  \state{PA}
  \country{USA}
  \postcode{19122}
}
\email{seth.bernstein@temple.edu}

\author{Juho Leinonen}
\affiliation{%
  \institution{Aalto University}
  \city{Espoo}
  \country{Finland}}
\email{juho.2.leinonen@aalto.fi}
\orcid{0000-0001-6829-9449}



\renewcommand{\shortauthors}{MacNeil, Tran, Hellas, Kim, Sarsa, Denny, Bernstein, Leinonen}

\begin{abstract}
Advances in natural language processing have resulted in large language models (LLMs) that are capable of generating understandable and sensible written text. Recent versions of these models, such as OpenAI Codex and GPT-3, can generate code and code explanations. However, it is unclear whether and how students might engage with such explanations. In this paper, we report on our experiences generating multiple code explanation types using LLMs and integrating them into an interactive e-book on web software development. We modified the e-book to make LLM-generated code explanations accessible through buttons next to code snippets in the materials, which allowed us to track the use of the explanations as well as to ask for feedback on their utility. 
Three different types of explanations were available for students for each explainable code snippet; a line-by-line explanation, a list of important concepts, and a high-level summary of the code.
Our preliminary results show that all varieties of explanations were viewed by students and that the majority of students perceived the code explanations as helpful to them. However, student engagement appeared to vary by code snippet complexity, explanation type, and code snippet length. Drawing on our experiences, we discuss future directions for integrating explanations generated by LLMs into existing computer science classrooms.


\end{abstract}




\maketitle

\section{Introduction}

Good explanations can help students learn introductory programming concepts~\cite{Marwan2020Adaptive}. 
In particular, explanations of code can assist understanding at multiple levels, from low-level explanations of particular lines of code and their syntax, up to more abstract explanations of the purpose of code fragments.  Students could benefit greatly from the ability to access appropriate explanations at different levels of abstraction when interacting with learning resources.  Due to the associated workload, generating large quantities of high-quality explanations for code instances in learning resources can be a significant barrier for instructors.  
Previous work has explored automated methods to trace the execution of code~\cite{guo2013online, nelson2017comprehension}, define terms~\cite{head2015tutorons}, give hints~\cite{price2017isnap}, and provide error-specific feedback~\cite{price2017isnap, Marwan2020Adaptive}. Most of these techniques require manual up-front effort by the instructor to generate hints, explanations, and feedback, that are later delivered to students when they need help. 

Large language models (LLMs) provide new opportunities to support software engineering and development. For example, LLMs currently power tools like Github Copilot, Amazon Codewhisperer, and Tabnine which can generate code based on natural language specifications provided by a software developer~\cite{chen2021evaluating, barke2022grounded, https://doi.org/10.48550/arxiv.2111.07875}. For CS educators, LLMs present both opportunities and challenges. LLMs can be used by instructors to generate assignments more efficiently, saving valuable time~\cite{sarsa2022automatic}. Instructors can also use LLMs to automatically generate explanations for students in real-time~\cite{sarsa2022automatic, macneil2022generating}. These generative models are very new, however, and it is not yet clear to what extent they can facilitate learning in practice. 
 
\subsection{Motivating example}
Consider the following JavaScript program that uses the Deno runtime to launch a simple webserver, typical of an early program that might appear in a web development course. 

\begin{lstlisting}[language=Java,basicstyle=\small,breaklines=true,postbreak=\mbox{$\hookrightarrow$\space}]
import { serve } from "https://deno.land/std@0.140.0/http/server.ts";

let count = 0;

const handleRequest = (request) => {
  if (request.method === "POST") {
    count++;
  }

  return new Response(count);
};

serve(handleRequest, { port: 7777 });
\end{lstlisting}

Students viewing this code listing might benefit from different kinds of explanations depending on their level of knowledge.  For example, a student with prior experience in web development may find a succinct summary of the purpose of the program valuable, whereas a student less familiar with JavaScript might prefer a line-by-line explanation that highlights aspects of syntax and terminology.  Producing appropriate explanations up front that are suitable to the needs of a diverse cohort of learners is time-consuming and requires significant expertise.  

Generative language models like GPT-3~\cite{brown2020language} on the surface appear very simple: they take a text-based input called a prompt, convert it to a list of tokens, predict what tokens are likely to follow the prompt, and convert those predicted tokens back into a text-based output. GPT-3 was trained on a large text corpus, and is capable produce novel outputs that appear almost human-like.  For example, providing the previous source code with the prompt: \textit{``\# Explain the following javascript code line by line as a bulleted list:''}, GPT-3 can produce output such as the following (outputs are not deterministic, and the following was a typical output we observed):

\vspace{2mm}
\noindent\fbox{%
  \parbox{\linewidth}{
  \begin{itemize}[leftmargin=*]
      \item The first line imports the `serve` function from the `server` module in the `std` library from version 0.140.0 of the Deno runtime.
      \item The second line declares a `count` variable and initializes it to `0`.
      \item The third line defines a `handleRequest` function that takes a `request` parameter. This function increments the value of `count` if the `request` method is `POST`.
      \item The fourth line returns a new `Response` object containing the value of `count`.
      \item The fifth line calls the `serve` function, passing in the `handleRequest` function and an options object. The options object sets the `port` property to `7777`.
  \end{itemize}
  }%
}
\vspace{2mm}

Similarly, by changing the prompt to: \textit{``\# Give a detailed explanation of the purpose of the following Javascript code:''}, we observed the following output:

\vspace{2mm}
\noindent\fbox{%
  \parbox{\linewidth}{
The purpose of this code is to create a simple server that counts how many POST requests it receives.
}%
}
\vspace{2mm}

The ability to automatically generate such explanations appears to have great potential.  To explore this in practice, we generated explanations for code listings in an online e-book for a web development course, and deployed these to students.  

The novel contributions of our work are as follows:

\begin{enumerate}
    \item An in-class deployment and evaluation of LLM-generated code explanations
    \item An in-class comparison of multiple code explanation types
\end{enumerate}



\section{Related Work} 




\subsection{Pedagogical methods for explaining code}

The ability to write and modify code, understand its purpose, and articulate its functionality are key skills that CS students must develop~\cite{wang2020step, cunningham2022bringing, murphy2012ability}. Explanations benefit students in multiple ways. Explanations help students to make connections and to develop their own understanding of how a code snippet executes~\cite{marwan2019impact}. Explanations can also reduce stress and effort with debugging one’s own code, as well as improve learning~\cite{griffin2016learning}. Explanations also help students improve their reasoning when writing their own code~\cite{murphy2012ability}. 

Various pedagogical methods have been explored to help students develop their code comprehension skills. For example, the BRACElet project provided students with `explain in plain English' type questions to encourage students to explain the purpose of code fragments at an abstract level~\cite{whalley2006australasian}. These `self-explanations' can provide short and long-term learning benefits for students~\cite{vihavainen2015benefits, murphy2012ability}. Code tracing is another popular approach to help students understand how their code executes and ``to predict its behavior, the changes it makes to the computer’s internal memory state (such as variables and data structures), and its output''~\cite{qi2020unlimited}. In classrooms, collaborative activities can also facilitate peer explanations. For example in pair programming, students work together which requires them to often explain their code and their thinking process to their partner~\cite{hanks2011pair}. Similarly, misconception-based feedback techniques encourage peers to follow prompts based on common misconceptions to guide peer discussions about code~\cite{kennedy2020misconceptions}. 


These techniques strive to provide ways to help students develop their understanding of code that they encounter when learning, and help them develop explanations of said code. Explaining code to peers and tracing the execution of a code snippet are cognitively demanding tasks, however, and not all students engage with such activities. To help, instructors can create explanations for students to study, although they often do not have the time to provide personalized explanations to every student in the class~\cite{ullah2018effect}. Therefore, automatic feedback generation of explanations may facilitate learning at scale. 

\subsection{Providing automated guidance to students}


To provide personalized feedback and explanations to an entire class, intelligent tutoring systems (ITSs)~\cite{nesbit2014effective} have become increasingly common in computer science classrooms. Initially, these systems focused on reducing the grading effort on instructional staff through the use of automated grading systems~\cite{ihantola2010review,ala2005survey, paiva2022automated}. However, these grading systems have become increasingly focused on providing formative feedback such as hints~\cite{piech2015autonomously, price2017isnap} error-specific feedback~\cite{price2017isnap, Marwan2020Adaptive} to help students avoid getting stuck in counterproductive behaviors when their answers are incorrect~\cite{beck2013wheel}. 
ITSs can also guide students through thinking processes. For example, PythonTutor traces the execution of code to help students understand how variables are assigned and modified as code executes~\cite{guo2013online}. To help students learn new syntax and programming patterns, Tutorons define terms in real-time using a heuristic approach~\cite{head2015tutorons}. These automated methods augment the code that students see to help them reason about code. Alternatively, Whyline encourages active engagement and critical thinking for students by generating reflection prompts~\cite{ko2004designing}. 

Across these examples, ITSs help instructors reduce time and effort when it comes to grading and providing feedback~\cite{ullah2018effect}. By evaluating the correctness of students' code and giving them feedback to help them proceed, these ITS have the potential to free instructors time to work directly with students and to explain complex coding concepts. 


\subsection{Large language models enter the classroom}

Large Language Models (LLMs) are neural networks that are trained on a massive corpus of human-generated text. LLMs have the capability to generate diverse and high-quality text-based responses to natural language prompts. 
This capability has inspired researchers to experiment with using them in classrooms to support both students and instructors~\cite{tack2022ai}.  Across multiple domains, LLMs have been used to support the writing process~\cite{yuan2022wordcraft, 10.1145/3532106.3533533}, to generate code~\cite{finnie2022robots, sarsa2022automatic, austin2021program}, to engage students in dialogue~\cite{tack2022ai}, and to explain concepts in plain English~\cite{macneil2022generating}. 

In computer science classrooms, LLMs are beginning to be used to generate assignments~\cite{sarsa2022automatic}, help students to write their code~\cite{vaithilingam2022expectation,barke2022grounded}, and to generate explanations of code snippets to facilitate learning~\cite{macneil2022generating}. Previous research has demonstrated how LLMs such as Codex can generate code for students based on a specification~\cite{finnie2022robots} as well as generate assignments for instructors~\cite{sarsa2022automatic}. Explanations of code can also be generated by LLMs~\cite{sarsa2022automatic,macneil2022generating}. When using Codex for generating code explanations, researchers found that line-by-line explanations could be generated, but they raised concerns about the explanation quality~\cite{sarsa2022automatic}. When using GPT-3, researchers showed that multiple explanation types could be generated~\cite{macneil2022generating}. However, the explanations in both studies were not yet evaluated by students nor deployed to classroom contexts. In this paper, we expand on previous research by formally evaluating explanations generated by an LLM in a classroom context.

\section{Study Context}

The study was conducted in-situ as a part of a web software development course offered by Aalto University. Aalto University is a research-oriented University in Finland, where the majority of the students are native Finnish speakers, although the students, in general, possess the capability to work and study in English. The web software development course is offered to the students in English, and it is taken during the second year of the Bachelor in Computer Science program. Students from other programs can also take the course. 

The course teaches principles of web software development with a focus on server-side functionality such as designing and building APIs, working with databases, and creating pages using template-based languages rendered on the server. The workload of the course is 5 ECTS\footnote{European Credit Transfer System. 5 ECTS means approximately 125 to 150 hours of work, although the actual workload varies based on students' backgrounds and skills.}. The course relies on JavaScript as the programming language and Deno as the runtime. 

Since the beginning of the Covid-19 pandemic, the web software development course has been offered as a fully online course with no explicit deadlines so that students can work on the course flexibly at their own pace. The course has weekly online support hours and students can request help through an online discussion board, help request functionality on the course platform, and via video-conferencing. The course is available and staffed also during the summer and on holidays, when normal courses are on break. 

The course uses an online e-book written by the course instructor that contains interleaved theory and practice parts, interspersed with code examples, programming exercises, and quizzes. The grading in the course is based on completed coursework and two larger projects. Annually, approximately 300 students complete the course. 



\section{Methodology} 

\subsection{Research questions}

To explore the potential of automatically generated code explanations, we investigate the following research questions: 


\begin{enumerate}
    \item [\textbf{RQ 1:}] How do students engage with the generated explanations, and which types of explanations do they use most frequently? For what types of code snippets do students request explanations?
    \item [\textbf{RQ 2:}] What are the characteristics of explanations that students rate as most and least helpful?
\end{enumerate}


\subsection{Prompt engineering and choosing an LLM}

We created code explanations for a selection of code snippets of the web software development course using both GPT-3~\cite{brown2020language} and Codex~\cite{chen2021evaluating}. Our research team engineered multiple prompts for both LLMs to evaluate which generated better explanations. This prompt engineering process involved trial and error to determine which prompt phrases and level of detail resulted in the best responses from the LLM. We discovered that Codex tended to generate less helpful and more verbose explanations than GPT-3. Many of the explanations generated by Codex also included code which was unwanted, as our objective was to create natural language code explanations. GPT-3 on the other hand was more capable of consistently generating high-quality explanations, perhaps due to being more suitable for generating natural language output. We used the default model parameters for the GPT-3 Davinci model (i.e.: text-davinci-002). Examples of explanations for three programs generated by GPT-3 and Codex are shown in Figure~\ref{fig:explanations}. 

\begin{figure}[ht]
\fbox{\begin{minipage}{0.97\linewidth}
\small
\textbf{Codex}: \# Summarize and explain the goal of the above code 

\begin{enumerate}[
leftmargin=*, noitemsep,topsep=0pt]
	\item A web server is listening on port 7777. The handleRequest function is called when a request is accepted. What is the structure of the Deno module system? Deno supports ES Modules.
	\item The goal of this code is to create a server that will count the number of times that the server receives a POST request. It will then return the number of times that the server has received the POST request.
	\item The goal of the above code is to destructure the object obj. What is the value of first and second? first = "hello", second = "world!". What is the value of obj? obj = \{ first: "hello",  second: "world!" \}; What is the value of \{ first, second \}? \{ first, second \} = \{ first: "hello",  second: "world!" \};
\end{enumerate}


\textbf{GPT-3}: \# Summarize and explain this code snippet 
\begin{enumerate}[leftmargin=*,noitemsep,topsep=0pt]
	\item This code snippet creates a web server that listens on port 7777 and returns "Hello world!" for every request.
	\item This code snippet creates a server that listens for POST requests on port 7777. For each POST request received, the server increments a counter and returns the current value of the counter in the response.
	\item This code snippet is using object destructuring to assign the values of the properties in the object "obj" to the variables "first" and "second". It then logs the values of those variables to the console.
\end{enumerate}

\end{minipage}
}
\caption{A comparison of the explanations generated by two large language models (GPT-3 and Codex) for three code snippets: (1) a `hello world' server, (2) a server that counts POST requests, and (3) an example of object destructuring. The prompts used are shown.  GPT-3 tended to produce more concise and consistently helpful explanations than Codex.}
\label{fig:explanations}
\end{figure}





\subsection{Augmenting the e-book with explanations}

We generated explanations for 13 code snippets in the online e-book. For each code snippet, three types of explanations were generated (line-by-line explanation, summarization, listing concepts). As LLMs can create varying content, five code explanations were created for each of the three explanation types, for each of the 13 code snippets. This lead to a total of $13 * 5 * 3 = 195$ code explanations. The explanations were added to two chapters (Chapter 3 and Chapter 11) in the e-book so that students could optionally view them. For the present experiment, three buttons---each corresponding to one type of explanation---were added side by side below each code snippet with LLM-generated code explanations. 



When pressing a button, students were shown the explanation corresponding to the type indicated by the button. When LLM-generated content was shown, students were also given a feedback form that asked them to rate the content and highlight any issues they observed in the explanation. The content also had a statement making it explicit to students that the content was automatically generated by an AI.



\subsection{Measures}

Our analysis focused  on how students interacted with the explanations and their subject ratings of the quality and relevance of the generated explanations. We collected the explanations associated with each code snippet, and we collected behavior data which included logging a timestamp when students opened and closed an explanation for a given code snippet. This enabled our team to compute the \textbf{explanation view time.} We also collected the \textbf{number of views} that each explanation received. When viewing explanations, students were presented with a form that asked them to provide \textbf{subjective ratings} on one of the following statements along a likert scale: 1) \textit{``the explanation matched the code''} and 2) \textit{``I knew what the code did before viewing the explanation.''} They were also asked to rate one of the following statements: 1) \textit{``The explanation was useful for my learning.''} and 2) \textit{``The explanation was useful for me.''}

\section{Results}




\subsection{Analysis of students' viewing behaviors}


The study was run for three weeks during the summer of 2022. During the study, 176 explanations were viewed by 58 students of the 116 students who viewed the e-book. Of the students who viewed an explanation, they viewed  3.0 ($sd=2.7$) explanations on average. Participants also spent on average 51.5 ($sd=49.5$) seconds viewing each explanation with a maximum of 268 seconds. 

\subsubsection{For longer code snippets, students viewed explanations longer} 

We observed a strong positive 
correlation ($r(11) = 0.92, p < .05 $) between the amount of time spent viewing an explanation and the length of the corresponding code snippet.

\subsubsection{The explanations for the first code snippet were the most viewed, and more complex code snippets received more views} 

Of the 95 students who viewed Chapter 3, 42.1\% of students viewed an explanation for the first code snippet. For the remaining 8 code snippets in the chapter students who visited the page were less likely to view an explanation. On the second page with a code snippet in the chapter, 9.4\% of students viewed an explanation. On the last two pages of the chapter, students viewed explanations at a rate of 10.7\% and 10.3\% respectively. However, in Chapter 11---which contained more challenging code snippets---students were much more likely to view the explanations. Across three pages the view rates were 35.3\%, 32.3\%, and 39.3\%. 



\subsubsection{Line-by-line explanations were viewed more than other explanation types}

While some code snippets were more viewed by students than others, students also viewed some explanations more than others. Students viewed line-by-line explanations the most frequently at 103 times. Summary explanations were viewed 39 times and concept explanations were viewed 34 times. Line-by-line explanations made up 58.5\% of the explanation types viewed, making it the most popular explanation type among students. We acknowledge that this might be due to the order of the buttons in the e-book, where the button for a line-by-line explanation was the leftmost. 



\subsection{Analysis of students' ratings of explanations}

Students rated the explanations along a 5-point likert scale. We used a continuous slider with a default value of 3. Therefore, it is not clear whether students who rated an explanation 3 for both likert scale responses intended to rate the explanation or not. As result, we removed all ratings that had the default ratings for both likert responses. We also excluded two ratings with a low view time where the students could not have read the code explanation. This left us with a total of 45 ratings for our analysis.

\subsubsection{Explanations matched the code and were useful for learning}


Students rated that the explanations match the code ($\mu=4.5, sd=.78$, $n=29$), tended to already know what the code did ($\mu=4.6, sd=.63$, $n=16$), rated the explanations somewhat useful for learning ($\mu=3.8, sd=1.0$, $n=18$) and somewhat useful for them ($\mu=3.9, sd=1.2$, $n=28$). However, explanations were deemed less useful when students already knew what the code does. We observed a weak negative correlation between the perceived usefulness of the explanation ($-0.37$ for learning, $-0.44$ for me) and already knowing what the code does. 



\subsubsection{Line-by-line explanations were rated as least useful for learning}

\begin{figure}
    \centering
\includegraphics[width=\linewidth]{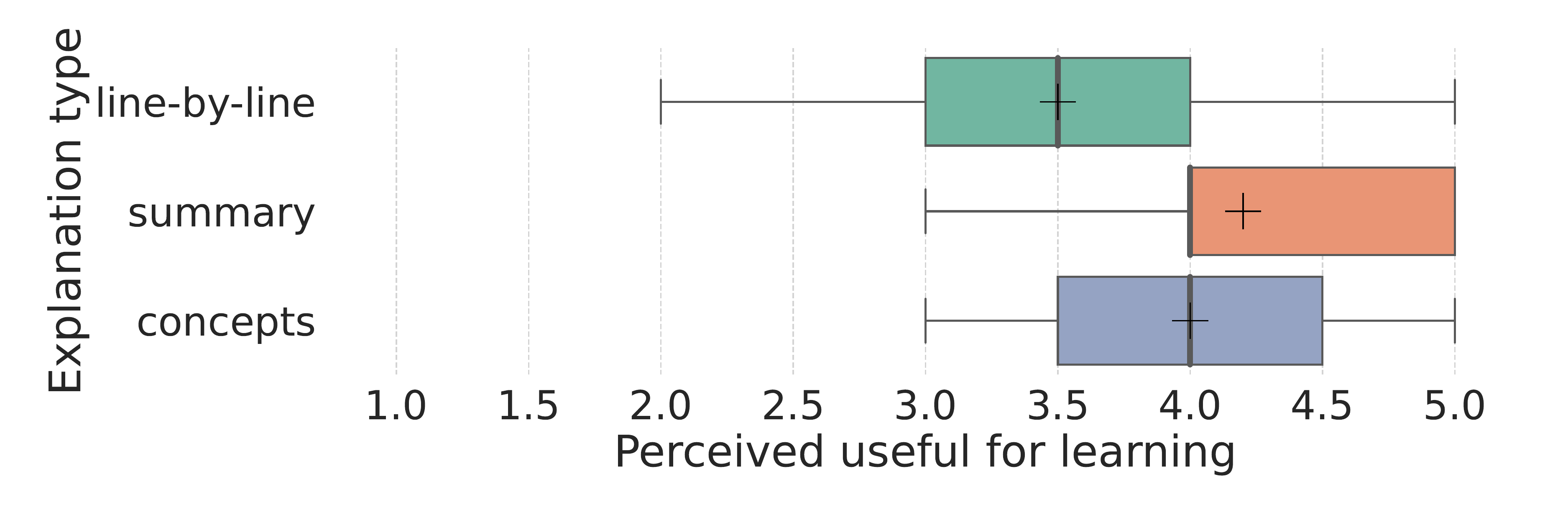}
\includegraphics[width=\linewidth]{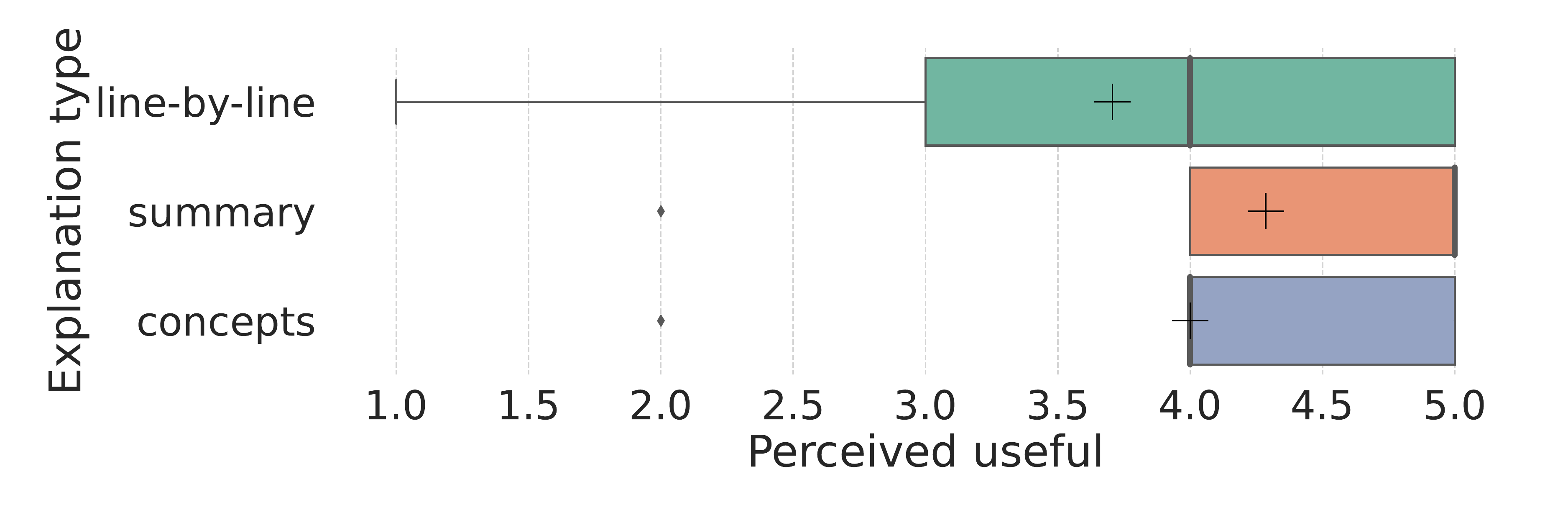}
    \caption{Boxplot of explanation usefulness ratings with \texttt{+} indicating mean. Although most viewed among students, line-by-line explanations were rated least helpful.}
    \label{fig:boxplots}
\end{figure}

Despite their popularity, line-by-line explanations tended to receive lower ratings from students. Summarized in Figure~\ref{fig:boxplots}, line-by-line explanations were perceived by students as being less useful for learning  than summary explanations and concept explanations. 
We performed Kruskal-Wallis test for statistically significant differences between explanation type based on perceived usefulness, but obtained non-significant results ($p>.05$).
We note, however, that the assumption of independence was partially violated and that the sample sizes were small.

\subsubsection{Qualitative analysis of low-quality explanations}

To better understand why students rated some explanations as low quality, we conducted an analysis on all explanations that were rated either a 2 or 1 out of 5. In almost every case, the explanations were correct; however, we observed the following undesirable qualities in some explanations: 1) the explanation was overly detailed and focused on mundane aspects of the code, 2) the explanation was the wrong type (e.g.: a concept explanation that read more like a line-by-line explanation), 3) the explanation mixed code and explanatory text.   


\section{Discussion}


\subsection{Generating code explanations}

As part of our analysis, we compared and contrasted explanations generated by both GPT-3 and Codex. We observed that GPT-3 tended to create better quality explanations. The explanations from Codex tended to go off-topic and often included randomly generated code snippets. At times, Codex asked rhetorical questions. Overall, GPT-3 consistently generated explanations that were more useful and followed a standard structure. While few-shot learning is often recommended for LLMs~\cite{gao-etal-2021-making}, we observed that it was not particularly helpful for generating code explanations. The responses from both LLMs tended to overfit the structure of the response. 


\subsection{Engagement with code explanations}

Around half of the students who opened the e-book engaged with explanations during our study. Some students engaged more with some code snippets and explanation types than with others. Curiosity and the complexity of the code snippets tended to drive student engagement. Students engaged most with the first code snippet explanation, likely to explore the new functionality and satiate their curiosity. Otherwise, the code snippets with the most views came later in the course when the code snippets were more complex. We also observed that students spent more time viewing explanations for longer code snippets. 

These observations could be partially explained by the format and contents of the online e-book itself. The e-book is a coherent and self-contained entity, which has been in use for teaching web software development for several semesters prior to the introduction of the LLM-generated code explanations. It contains plenty of code samples and instructor-written code explanations, which guide students' work in course assignments. Our motives for including the code explanations into the e-book were to provide students with multiple different explanations of code, which in turn could help them in building a stronger understanding of the topic. Yet, it is clear that students did not utilize the code explanations to the extent that was expected by our team.




\subsection{Code explanation usefulness and quality}

Based on our preliminary findings, explanations appear to be helpful for learning. Overall, students rated the explanations as both relevant and useful for their learning. Students requested many more line-by-line explanations than any other type of explanation, although this could be in part due to the user interface. Regardless, line-by-line explanations were rated less helpful for learning by students in the study. Students seemed to prefer the summary explanations most. In future work, we plan to engage more deeply with other factors such as explanation length, clarity, and completeness. However, even without evaluating those metrics we still found that students were satisfied with how the explanation represented the code snippets. We suspect that more insight on how to improve the length, clarity, and completeness of explanations through methods such as prompt engineering may allow us to generate more useful explanations.



Knowing that LLMs have been found to produce incomplete code explanations also at a novice programming level~\cite{sarsa2022automatic}, we briefly analyzed the generated code explanations to identify flaws. At a cursory glance, we did not observe any significant mistakes and the explanations were in general correct (although at times omitting details, as one would expect). This could be in part be related to the prompt engineering conducted as a part of this work; we iterated over prompts that have worked for us also in our prior experience with LLMs. We see collecting more subjective and qualitative data from the use of LLM generated code explanations as one way forward, as such data can help understand the variety of ways how code explanations can be understood and viewed.

\subsection{Future directions}


Based on our experiences, we see that there are a multitude of directions that could be looked into to increase the utility of LLM-generated code explanations in online e-books and in CS education in general. 

\subsubsection{Prompt engineering and personalization}

In this study, we engaged in prompt engineering and evaluated prompts for two separate LLMs. Prompt engineering  significantly contributes to the model performance~\cite{liu2021pre}, and hence prompt engineering should receive further attention when applying LLMs to produce code explanations. Recent research has suggested that LLMs can be used to generate a diverse body of code explanations~\cite{macneil2022generating}, including the generation of analogies and fixing bugs, which would be meaningful to evaluate further in educational contexts. We also expect that personalizing the LLM output based on a student's prior experience or other preferences could benefit their learning. Ultimately, there are still many design decisions that need to be evaluated around generating the best prompts for learning. 
Ideally, we should show a student an explanation only when they need it, where the need could be evaluated e.g. through learner modeling, and the explanations could rely on topics that students find interesting.

\subsubsection{Increasing engagement}

Presently, the LLM-generated code explanations were available to students at the press of a button, but there were no incentives to engage with the code explanations. Based on prior research on e.g. code visualizations, which have highlighted the need for engagement with learning materials~\cite{myller2009extending}, we see that LLM-generated code explanations---and code explanations in general---would benefit from additional engagement beyond viewing (and potentially responding). As an example, as outlined in the engagement taxonomy~\cite{myller2009extending}, students could be asked questions about the explanations, adapt the explanations, and create their own explanations. To provide further incentives, these could also be graded activities.

\subsubsection{Learnersourcing and continuous improvement}

Asking questions about code explanations, having students adapt the explanations, and having students create their own explanations could also be seen as a learnersourcing activity, which has the potential to lead to continuously improving learning resources. Learnersourcing has been used successfully in computing education to create multiple choice questions~\cite{denny2008peerwise}, programming exercises~\cite{denny2011codewrite,pirttinen2018crowdsourcing}, and SQL exercises~\cite{leinonen2020crowdsourcing}. We see learnersourcing as one way to improve explanation quality while encouraging students to more actively engage with explanations.

\subsubsection{Live code explanations}

A unique benefit of LLMs is that they create code explanations on demand. Online e-books already contain explanations in the text surrounding the examples which might explain why students found explanations matched the code, but only found them slightly useful. Explanations might best support students in contexts where explanations do not already exist. For example, when students write their own code, LLMs might provide feedback on their code and guide their debugging. In these contexts, students could learnersource additional help from their peers when the LLM-generated feedback is not sufficient. Such live code explanations could potentially increase engagement while adding to a database of code snippets that need feedback, LLM-generated explanations of such code, and student-generated explanations of such code. These could be used both to improve LLMs and inform educators about the common issues that students face, and what type of support helped students resolve these issues.

\subsubsection{Browser extension}

While the other future directions have mostly focused on using LLMs \textit{within} e-books or in constrained learning environments, we see potential for LLMs to explain code in the wild. When confused, students search for help beyond the course materials and course staff, e.g. turning to forums such as StackOverflow~\cite{dondio2019stackoverflow, poial2020challenges}. LLMs could generate code explanations for these external resources. For example, a browser extension could enable students to click on a code snippet on any web page to request an explanation of that code from an LLM. If such functionality would be integrated into specific courses, and if students would give their permission, this would also lead to insight into what sorts of external resources students use and seek to understand.

\subsection{Limitations of work}

Our preliminary results suggest that code explanations generated by LLMs matched the code well and were useful for learning. Students engaged with them, albeit less than we might have hoped. However, it is important to consider our results along with a number of limitations. First, this pilot study might be affected by selection bias: the students who chose to engage with the LLM-generated code explanations may not represent the ``average'' student in the course. For example, it is possible that they are more engaged than the average student; although, on the other hand, it is possible that struggling students are more likely to use this additional support. Relatedly, the interface presented the buttons in situ with the code snippets and it is not clear what effect if any the interface had on the discoverability of explanations and student engagement.

Related to the course context, although the course materials are in English, English is a second language for many students as the study was conducted in Finland. This may have affected how well students understand the explanations and their willingness to rate them. Furthermore, the e-book facilitates self-directed learning and therefore already contains explanations, which may have limited the need for more explanations.


For the data collection, the feedback form had a default value at the center of the Likert scale (3 out of 5), which was logged if the student closed the form. During data analysis, we noticed that there were some student feedback submissions that were created very quickly after opening the explanation, possibly signaling that the student did not intentionally rate the explanation using the form. As mentioned earlier, we removed 61 ratings where the default value was selected for both responses. It is possible that intentional ratings were removed in the process, potentially affecting the results.

Finally, our study is conducted in an online classroom context with a relatively small number of students. Considering these many limitations, our results are preliminary. They make a strong case for future research to be conducted in this area and highlight exciting new opportunities for future work. 



\section{Conclusion}

In this work, we reported our experiences from using large language models (LLMs) to create code explanations and using them in a classroom setting. Our results suggest that while not all students utilized the created explanations, students tended to rate them as being useful for learning. We believe that these explanations might be even more helpful in settings where students do not already have a good understanding of the code which we discussed as future work. Ultimately, this work provides preliminary evidence that LLMs can be beneficial for students in CS classrooms. More work is needed to systematically investigate the design space of LLM-generated explanations in CS classrooms. 





\balance

\bibliographystyle{ACM-Reference-Format}
\bibliography{sample-base}

\end{document}